\documentclass[a4paper,11pt]{article}
\pdfoutput=1 

\usepackage{jheppub} 

\usepackage[T1]{fontenc}
\usepackage{multirow}
\usepackage{amsmath,amssymb,relsize,slashed}
\usepackage[normalem]{ulem}

\allowdisplaybreaks
 
\definecolor{nicered}{rgb}{0.7,0.1,0.1}
\definecolor{nicegreen}{rgb}{0.1,0.5,0.1}
\definecolor{niceblue}{rgb}{0.0,0.1,0.7}
\hypersetup{colorlinks,citecolor=niceblue,linkcolor=niceblue,urlcolor=niceblue}

\def \beq{\begin{equation}}
\def \eeq{\end{equation}}
\def \bea{\begin{eqnarray}}
\def \eea{\end{eqnarray}}

\title{Exact two-loop amplitudes for Higgs plus jet production with a cubic Higgs self-coupling}

\author[a]{Ulrich Haisch}
\author[a]{and Marco Niggetiedt}

\affiliation[a]{Max Planck Institute for Physics, \\ Boltzmannstr.~8, 85748 Garching, Germany}

\emailAdd{haisch@mpp.mpg.de}
\emailAdd{marco.niggetiedt@mpp.mpg.de}

\preprint{MPP-2024-197}

\abstract{We compute the corrections to the two-loop amplitudes for $gg \to h$, $gg \to hg$, $qg \to hq$, and $q \bar q \to hg$ due to a modified cubic Higgs self-coupling. The exact dependence on the Higgs and top-quark masses across the entire $2 \to 2$ phase space is determined by numerically solving a system of differential equations for the relevant master integrals. The~calculated amplitudes are crucial for evaluating the impact of the considered corrections on exclusive~$pp \to hj$ production at the hadronic event level. As an application, we calculate the non-universal, kinematic-dependent cubic Higgs self-coupling corrections to the Higgs-boson transverse momentum distribution in gluon-gluon fusion Higgs production for arbitrary values of the transverse~momentum.}

\begin{document} 
\maketitle
\flushbottom

\section{Introduction}
\label{sec:introduction}

Investigating the scalar potential of the spin-0 state $h$ discovered by the ATLAS and CMS collaborations~\cite{ATLAS:2012yve,CMS:2012qbp} is a cornerstone of the Large Hadron Collider (LHC) program. The~first step in investigating the Higgs potential $V$ is to study the cubic Higgs self-coupling, which can be expressed in a model-independent manner as
\beq \label{eq:V}
V \supset \kappa_3 \hspace{0.25mm} \lambda \hspace{0.25mm} v \hspace{0.25mm} h^3 \,.
\eeq
Here, $\lambda = m_h^2/(2 \hspace{0.25mm} v^2) \simeq 0.13$, where $m_h \simeq 125 \, {\rm GeV}$ is the Higgs mass and $v \simeq 246 \, {\rm GeV}$ is the electroweak (EW) vacuum expectation value. Note that the normalization of the terms in~(\ref{eq:V}) is chosen so that in the Standard Model~(SM), $\kappa_3 = 1$. In the presence of physics beyond the SM, the coefficient $\kappa_3$ will typically differ from 1. For example, consider the term
\beq \label{eq:LSMEFT}
{\cal L}_{\rm SMEFT}\supset -\frac{\lambda \hspace{0.25mm} \bar c_6}{v^2} \hspace{0.25mm} \left |H \right |^6 \,,
\eeq
within the SM effective field theory~(SMEFT)~\cite{Buchmuller:1985jz,Grzadkowski:2010es,Brivio:2017vri,Isidori:2023pyp}, where $H$ represents the usual Higgs doublet. In this scenario, $\kappa_3$ is related to the Wilson coefficient $\bar c_6$, with their relationship at tree level given by:
\beq \label{eq:kappa3c6relation}
\kappa_3 = 1 + \bar c_6 \,.
\eeq

The most direct way to probe the parameter $\kappa_3$, or equivalently the Wilson coefficient~$\bar c_6$, at the LHC is through double-Higgs production via gluon-gluon fusion~(ggF), as a non-zero~$\bar c_6$ affects this process at leading order~(LO) in QCD. However, additional constraints on the Wilson coefficient $\bar c_6$ can also be obtained from single-Higgs production and decays at the LHC~\cite{Gorbahn:2016uoy,Degrassi:2016wml,Bizon:2016wgr,DiVita:2017eyz,Maltoni:2017ims,Maltoni:2018ttu,Gorbahn:2019lwq,Degrassi:2019yix,Haisch:2021hvy,ATLAS:2022jtk,Gao:2023bll,CMS:2024awa}, as these processes become sensitive to modifications of the cubic Higgs self-coupling when next-to-leading order~(NLO)~EW corrections are considered. These types of corrections can be categorized into two classes. First, corrections quadratic in $\bar c_6$ that are associated to Higgs wave-function renormalization~(WFR). The~WFR contributions are universal, leading to the same constant shift in all on-shell Higgs distributions. Second, there are corrections linear in $\bar c_6$ that contribute to single-Higgs production and decays in a process-dependent manner and are generally kinematic-dependent. At~the~differential level, the non-universal effects involving $\bar c_6$ have been exactly calculated for vector-boson fusion~$pp \to hjj~$\cite{Degrassi:2016wml,Bizon:2016wgr}, $pp \to Vh$~\cite{Degrassi:2016wml,Bizon:2016wgr}, $pp \to t \bar t h$~\cite{Degrassi:2016wml}, $pp \to t h j$~\cite{Maltoni:2017ims}, and off-shell Higgs production in $pp \to h^\ast \to ZZ \to 4 \ell$~\cite{Haisch:2021hvy}. However, for ggF production, only approximate results are currently available~\cite{Gorbahn:2019lwq,Gao:2023bll}, allowing for a reliable calculation of the differential $pp \to hj$ production cross section below the top-quark threshold, but in general not above~it.

The primary goal of our work is to close the latter gap by performing an exact numerical calculation of the non-universal ${\cal O} (\bar c_6)$ corrections to the two-loop amplitudes $gg \to hg$, $qg \to hq$, and $q \bar q \to hg$, which are essential for Higgs plus jet production. This will yield precise results for the squared matrix elements across the entire $2 \to 2$ phase space. The~basic steps of our calculation are outlined in~Section~\ref{sec:calculation}, where we also present formulas for the relevant squared matrix elements. These formulas are written in a form that allows for an easy implementation into a Monte~Carlo~(MC) code to simulate $pp \to hj$ production. In~Section~\ref{sec:numerics}, we apply the obtained formulas to compute the non-universal ${\cal O} (\bar c_6)$ corrections to the transverse momentum~($p_{T,h}$) distribution of the Higgs boson in ggF production for arbitrary values~of~$p_{T,h}$. We also compare our results with those from~\cite{Gao:2023bll}, which uses a heavy top-quark mass expansion (HTE) combined with a Pad{\'e} approximation to extend the convergence range of the HTE. Section~\ref{sec:conclusions} provides our conclusions and a perspective on possible future research~directions.

\section{Calculation}
\label{sec:calculation}

\begin{figure}[!t]
\begin{center}
\includegraphics[width=0.975\textwidth]{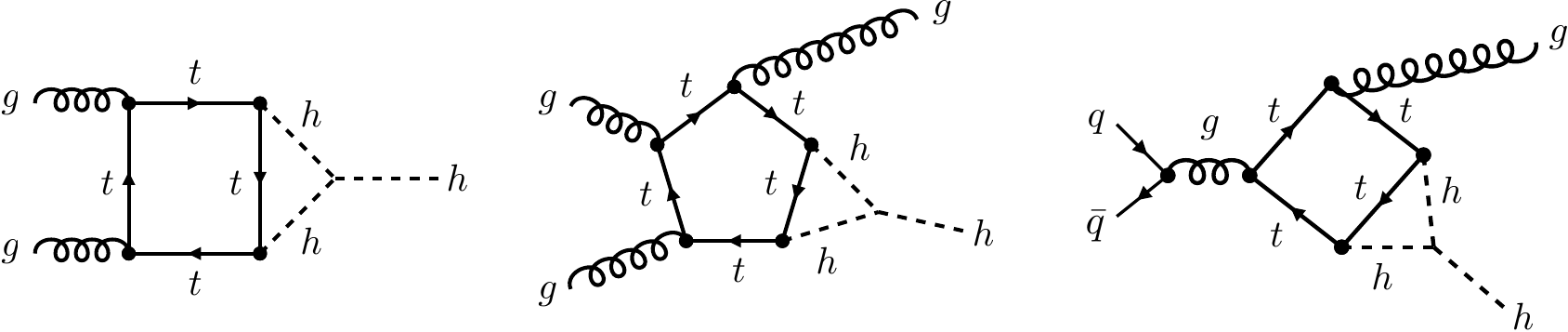} 
\vspace{2mm}
\caption{\label{fig:diagrams} Examples of two-loop Feynman diagrams with an insertion of a cubic Higgs self-coupling~(black square) that contribute to the $gg \to h$~(left), $gg \to hg$~(middle), and $q \bar q \to h g$~(right) channel, respectively.}
\end{center}
\end{figure}

Determining the precise Higgs and top-quark mass dependence of corrections to exclusive ggF Higgs production with a modified cubic Higgs self-coupling presents a challenge primarily due to the complex analytic structure of the two-loop amplitudes involved in the $pp \to hj$ process. This complexity arises from the intricate nature of the underlying Feynman integrals. Example diagrams illustrating this are shown in Figure~\ref{fig:diagrams}. From left to right, we display a representative contribution to the $gg \to h$, $gg \to hg$, and $q \bar q \to hg$ channels.

In this work, the relevant squared matrix elements have been computed using the methodology outlined and successfully applied in~\cite{Czakon:2020vql,Czakon:2021yub,Czakon:2023kqm,Niggetiedt:2024nmp,Czakon:2024ywb}. The two-loop amplitudes were reduced to linear combinations of master integrals using the software packages {\tt Kira} and {\tt FireFly}~\cite{Maierhofer:2017gsa,Maierhofer:2018gpa,Klappert:2019emp,Klappert:2020aqs,Klappert:2020nbg}. To~simplify intermediate algebraic manipulations, the ratio of the squared Higgs mass to the squared top-quark mass was fixed to $x = m_h^2/m_t^2 = 12/23$, which corresponds to $m_h \simeq 125 \, {\rm GeV}$ and $m_t \simeq 173 \, {\rm GeV}$. The same programs were also utilized to derive systems of first-order homogeneous linear differential equations~\cite{Kotikov:1990kg,Kotikov:1991pm,Kotikov:1991hm,Remiddi:1997ny} in the top-quark mass and the kinematic invariants, which are satisfied by the master integrals. To apply the differential equations, an initial condition is needed, which we obtain using a diagrammatic HTE~\cite{Gorishnii:1989dd,Smirnov:1990rz,Smirnov:1994tg,Smirnov:2002pj}. This approach also enables us to validate all the HTE results for the two-loop $pp \to hj$ amplitudes reported in~\cite{Gorbahn:2019lwq,Gao:2023bll}. 

Without imposing any restrictions on the masses and the $2 \to 2$ phase space, the parameter space for the relevant two-loop $pp \to hj$ amplitudes is described by three dimensionless variables, which are constructed from four independent dimensionful variables:
\beq \label{eq:parameterisation}
\rho = \frac{m_t^2}{\hat s} \,, \qquad \xi = 1 - \frac{m_h^2}{\hat s} \,, \qquad \chi = \frac{\hat t}{\hat t + \hat u} \,.
\eeq
Here, $\hat s$, $\hat t$, and $\hat u$ are the usual Mandelstam variables, which satisfy $\hat{s} + \hat{t} + \hat{u} = m_h^2$. The~two-dimensional subspace, parametrized by $\xi \in [0,1]$ and $\chi \in [0,1]$, is explored using the differential equations for the master integrals. Specifically, the differential equations in~$\rho$ are used to transport the initial condition from $\rho \to \infty$ to the two-dimensional plane at $x = 12/23$. The endpoint of this initial evolution then serves as a starting point for further evolutions in the $\xi\hspace{0.25mm}$--$\hspace{0.25mm}\chi$ plane, allowing the relevant squared matrix elements to be efficiently sampled over the whole $2 \to 2$ phase space. 

\begin{figure}[!t]
\begin{center}
\includegraphics[width=0.45\textwidth]{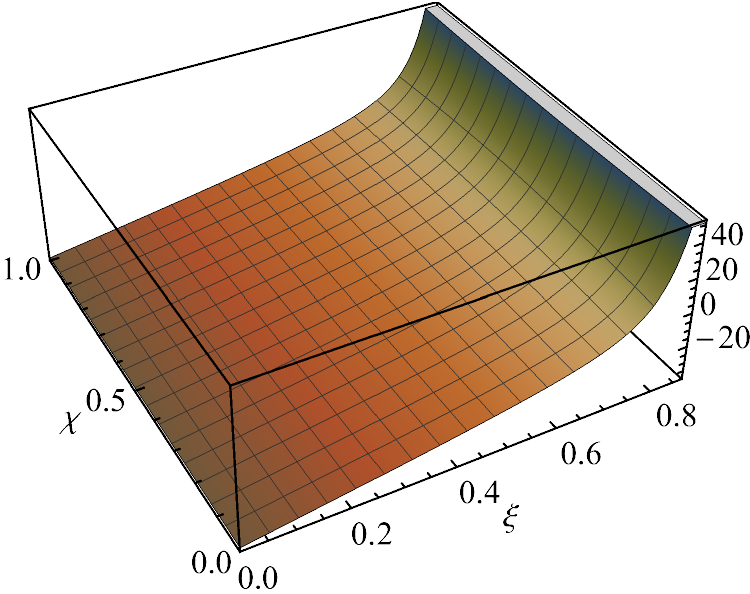} \qquad \includegraphics[width=0.45\textwidth]{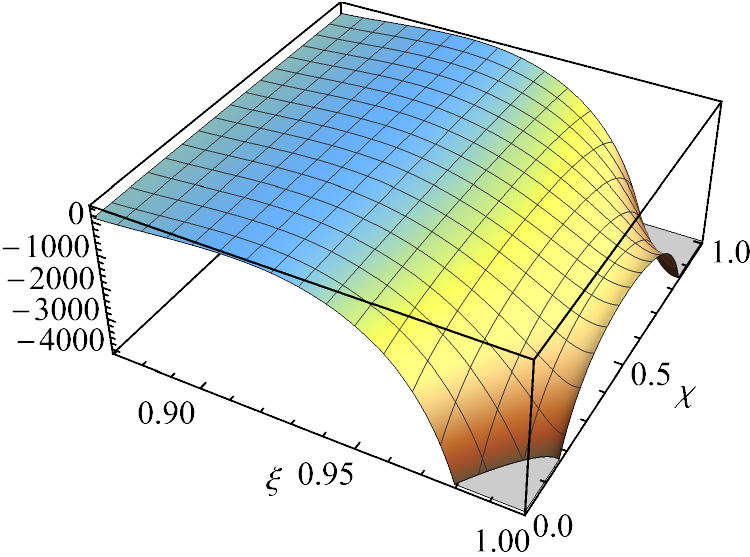} 

\vspace{4mm}

\includegraphics[width=0.45\textwidth]{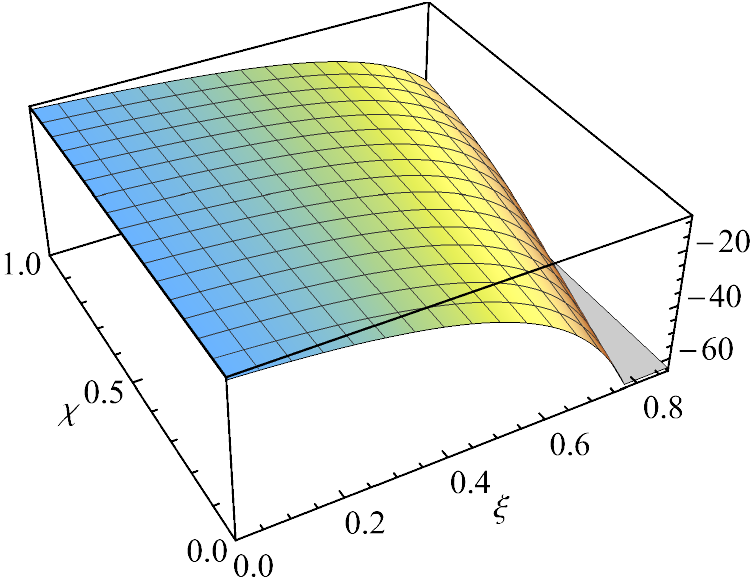} \qquad \includegraphics[width=0.45\textwidth]{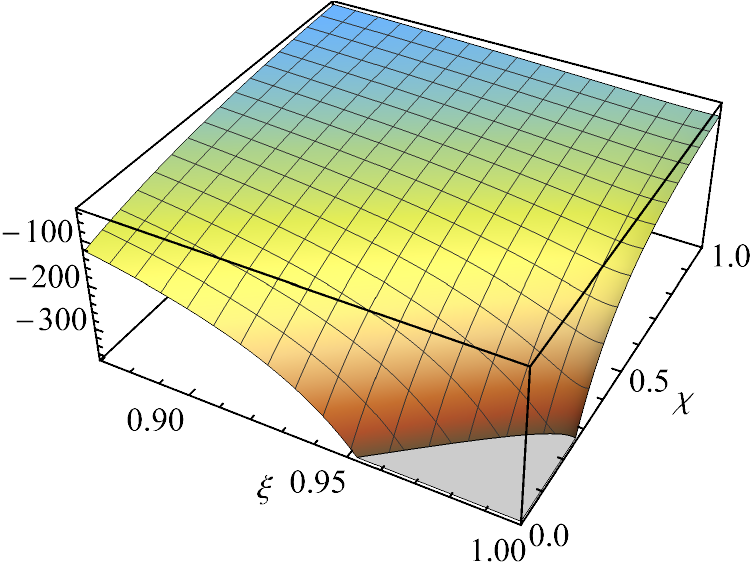} 
\vspace{4mm}

\includegraphics[width=0.45\textwidth]{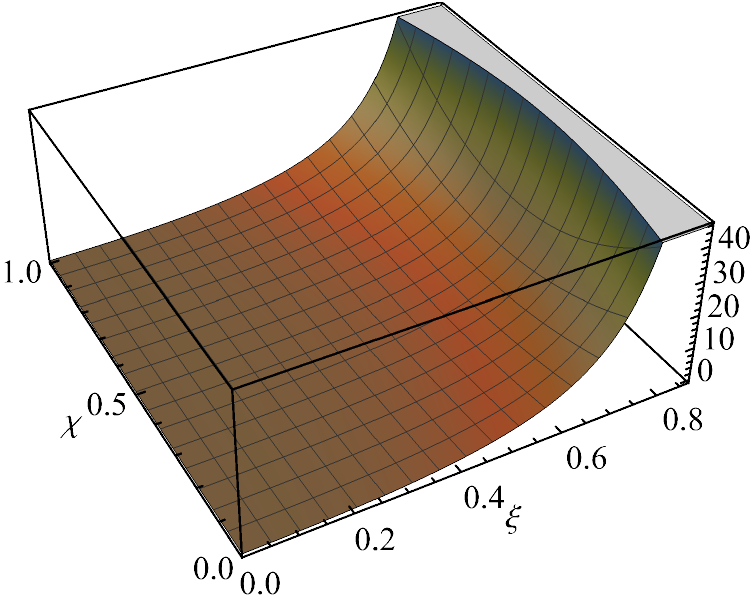} \qquad \includegraphics[width=0.45\textwidth]{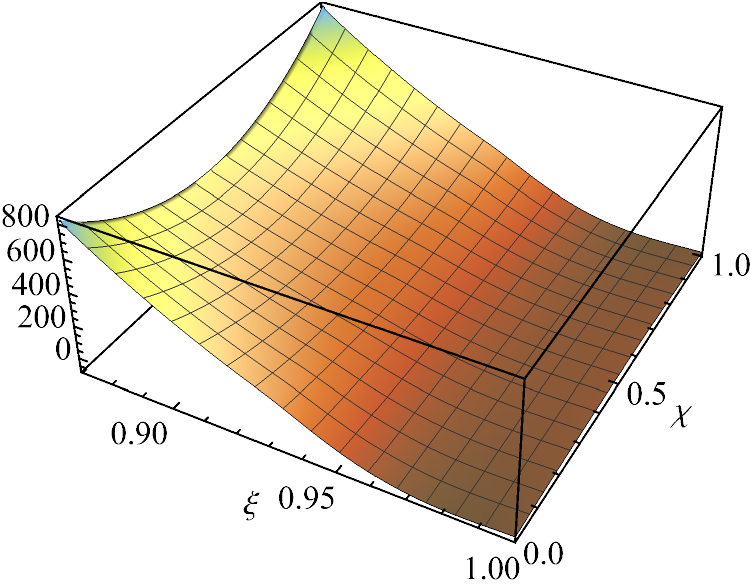} 
\vspace{6mm}
\caption{\label{fig:regulated} The regulated squared matrix elements $2 \hspace{0,25mm} {\rm Re} \hspace{0,25mm} \big \langle {\cal M}_{\rm exact}^{(0)} \big | {\cal M}_{\rm exact}^{(1)} \big \rangle_X \big |_{\rm regulated}$, as defined in~(\ref{eq:regulated}), are presented on the $z$-axis as a function of $\xi$ and $\chi$. Results for the $gg \to hg$~(first~row), $q g \to hq$~(second~row), and $q \bar q \to hg$~(third~row) process are shown, with the left column depicting the region below the threshold for intermediate top-quark pair production at $\hat s = 4 m_t^2$, and the right column showing the region above it. In each case, the appropriate normalization factor~$N_X$ given in~(\ref{eq:NX}) has been factored~out.}
\end{center}
\end{figure}

To enhance the convergence of the squared matrix elements near the phase space boundaries, where the amplitudes tend to diverge, we calculate a regulated version rather than computing the exact result directly, defined for a given partonic process $X$ as:
\bea \label{eq:regulated}
\big \langle {\cal M}_{\rm exact}^{(0)} \big | {\cal M}_{\rm exact}^{(1)} \big \rangle_X \big |_{\rm regulated} = \big \langle {\cal M}_{\rm exact}^{(0)} \big | {\cal M}_{\rm exact}^{(1)} \big \rangle_X - r_F \, \big \langle {\cal M}_{\rm HTL}^{(0)} \big | {\cal M}_{\rm HTL}^{(1)} \big \rangle_X \,. \hspace{6mm} 
\eea
Here, ${\cal M}_{\rm exact}^{(0)}$ and ${\cal M}_{\rm exact}^{(1)}$ represent the exact amplitudes for the LO QCD SM and NLO EW SMEFT corrections, respectively, while ${\cal M}_{\rm HTL}^{(0)}$ and ${\cal M}_{\rm HTL}^{(1)}$ denote the corresponding amplitudes in the limit of infinitely heavy top-quark mass~(HTL). The term $r_F$ appearing in~(\ref{eq:regulated}) is given for $x = 12/23$ by 
\beq \label{eq:forrmfactorratio}
r_F = \frac{{\rm Re} \hspace{0.5mm} \big \langle F_{\rm exact}^{(0)} \big | F_{\rm exact}^{(1) } \big \rangle}{\big \langle F_{\rm HTL}^{(0)} \big | F_{\rm HTL}^{(1)} \big \rangle} = 1.9166472020710834061 \,,
\eeq
where $F_{\rm exact}^{(0)}$, $F_{\rm exact}^{(1)}$, $F_{\rm HTL}^{(0)}$, and $F_{\rm HTL}^{(1)}$ denote the respective amplitudes of the $gg \to h$~process. The result~(\ref{eq:forrmfactorratio}) was achieved by performing a HTE of $F_{\rm exact}^{(1)}$ up to order $x^{100}$, which extends the depth of the HTEs obtained in~\cite{Degrassi:2016wml,Gorbahn:2019lwq,Gao:2023bll} by more than 90 terms. The last term in~(\ref{eq:regulated}) can be written as 
\beq \label{eq:HTL}
\begin{split}
\big \langle {\cal M}_{\rm HTL}^{(0)} \big | {\cal M}_{\rm HTL}^{(1)} \big \rangle_X & = N_X \hspace{0.125mm} H_X \,,
\end{split} 
\eeq
where the normalization factors $N_X$ are given in the case of the $gg \to hg$, $qg \to hq$, and $q \bar q \to hg$ channels by 
\beq \label{eq:NX}
N_{gg \to hg} = \frac{\alpha_s ^3}{32768 \hspace{0.125mm} \pi ^3} \hspace{0.5mm} \frac{\lambda \hspace{0.25mm} \bar c_6}{v^2} \hspace{0.5mm} \hat s \,, \quad 
N_{qg \to hq} = \frac{\alpha_s ^3}{12288 \hspace{0.125mm} \pi ^3} \hspace{0.5mm} \frac{\lambda \hspace{0.25mm} \bar c_6}{v^2} \hspace{0.5mm} \hat s \,, \quad 
N_{q \bar q \to hg} = \frac{\alpha_s ^3}{4608 \hspace{0.125mm} \pi ^3} \hspace{0.5mm} \frac{\lambda \hspace{0.25mm} \bar c_6}{v^2} \hspace{0.5mm} \hat s \,,
\eeq
with $\alpha_s = g_s^2/(4 \pi)$ denoting the strong coupling constant. In terms of the kinematic variables introduced in~(\ref{eq:parameterisation}), the corresponding coefficients~$H_X$ are 
\beq \label{eq:HX}
\begin{split}
H_{gg \to hg} & = \frac{128 \left( \xi^4 \left(\chi^2-\chi+1\right)^2 -2 \hspace{0.25mm} \xi^3+3 \hspace{0.25mm} \xi^2-2 \hspace{0.25mm} \xi+1\right)}{3 \hspace{0.25mm} \xi^2 \hspace{0.25mm} \chi \left (\chi-1 \right )} \, f \left (\frac{12}{23} \right) \,, \\[2mm]
H_{q g \to hq} & = -\frac{32 \left( \xi^2 \left ( \chi - 1 \right )^2 + 1 \right)}{9 \hspace{0.25mm} \xi \hspace{0.25mm} \chi} \, f \left (\frac{12}{23} \right) \,, \\[2mm] 
H_{q \bar q \to hg} & = -\frac{32 \hspace{0.25mm} \xi^2 \left(2 \hspace{0.25mm} \chi^2-2 \hspace{0.25mm} \chi+1\right)}{9} \, f \left (\frac{12}{23} \right) \,, 
\end{split} 
\eeq
with 
\beq \label{eq:fx}
f(x) = 12 \ln x + 4 \sqrt{3} \hspace{0.25mm} \pi - 23 \,. 
\eeq
 We emphasize that for a given channel, the matrix element ${\cal M}_{\rm HTL}^{(1)}$ is proportional to~${\cal M}_{\rm HTL}^{(0)}$. As a result, in the strict HTL, the kinematic distributions produced by the NLO~EW~SMEFT and LO~QCD~SM squared matrix elements for the processes $gg \to hg$, $qg \to hq$, and $q \bar q \to hg$ differ only by a constant multiplicative factor. This is a known feature~\cite{Gorbahn:2019lwq}, since, for dimensional reasons, the HTL terms can only arise from a single effective interaction of the form $|H|^2 \hspace{0.25mm} G_{\mu \nu}^a G^{a \hspace{0.25mm} \mu \nu}$, where $G_{\mu \nu}^a$ represents the QCD field strength tensor. Beyond LO effects in the HTE, however, break the proportionality between the NLO EW SMEFT and LO QCD SM squared matrix elements. Since these effects are in general numerically small, it follows that the dominant singularity structure of $\big \langle {\cal M}_{\rm exact}^{(0)} \big | {\cal M}_{\rm exact}^{(1)} \big \rangle_X$ matches that of $\big \langle {\cal M}_{\rm HTL}^{(0)} \big | {\cal M}_{\rm HTL}^{(1)} \big \rangle_X$, which in turn corresponds to the structure of $\big \langle {\cal M}_{\rm HTL}^{(0)} \big | {\cal M}_{\rm HTL}^{(0)} \big \rangle_X$. These observations explain why, in practice, computing the regulated version of the NLO EW SMEFT squared matrix elements, as done in~(\ref{eq:regulated}), offers greater numerical stability compared to directly calculating the exact results.

Our numerical results for the regulated squared matrix elements for the three relevant partonic processes are illustrated in Figure~\ref{fig:regulated}. The results below the threshold for intermediate top-quark pair production at $\hat s = 4 m_t^2$ are displayed on the left-hand side, while those above the threshold are shown on the right-hand side. We observe that, aside from the boundaries of the~$2 \to 2$ phase space, the regulated squared matrix elements $\big \langle {\cal M}_{\rm exact}^{(0)} \big | {\cal M}_{\rm exact}^{(1)} \big \rangle_X \big |_{\rm regulated}$ remain relatively flat across the $\xi\hspace{0.25mm}$--$\hspace{0.25mm}\chi$ plane. This suggests that the leading kinematic singularities of $\big \langle {\cal M}_{\rm exact}^{(0)} \big | {\cal M}_{\rm exact}^{(1)} \big \rangle_X$ are effectively captured by the subtraction term proportional to the~HTL~result $\big \langle {\cal M}_{\rm HTL}^{(0)} \big | {\cal M}_{\rm HTL}^{(1)} \big \rangle_X$. Using the HTE and the differential equations for numerical evolution in the $\xi\hspace{0.25mm}$--$\hspace{0.25mm}\chi$ plane, we generated dense two-dimensional grids for a fixed value of $x = 12/23$. These grids are well-suited for interpolating the relevant regulated squared matrix elements for the $gg \to hg$, $q g \to h q$, and $q \bar q \to hg$ channels. Each grid comprises approximately $7 \cdot 10^6$ high-precision samples of the regulated squared matrix elements. We expect that these grids can also be used in phenomenological applications for slightly different fixed ratios $x = m_h^2/m_t^2$, deviating by no more than $1\%$ from the true ratio. 

\section{Numerics}
\label{sec:numerics}

As a phenomenological application of the calculated two-dimensional grids for the squared matrix elements, we determine the magnitude of the non-universal ${\cal O} (\bar c_6)$ corrections to the~$p_{T,h}$~distribution in $pp \to hj$ production at the LHC in this section. At the level of squared matrix elements, we define the non-universal ${\cal O} (\bar c_6)$ corrections in a fully differential manner, meaning point-by-point across the whole $2 \to 2$ phase space, as
\beq \label{eq:C1}
\bar c_6 \hspace{0.5mm} C_1^X \left (\left \{p_n \right \} \right ) = \frac{2 \hspace{0,25mm} {\rm Re} \hspace{0,25mm} \big \langle {\cal M}_{\rm exact}^{(0)} \big | {\cal M}_{\rm exact}^{(1)} \big \rangle_X}{\big \langle {\cal M}_{\rm exact}^{(0)} \big | {\cal M}_{\rm exact}^{(0)} \big \rangle_X} \,, 
\eeq
where the numerator appearing on the right-hand side can be calculated from~(\ref{eq:regulated}). Note that in~(\ref{eq:C1}), we have explicitly indicated the dependence on the external momenta~$p_n$ in the Born configuration in parentheses, corresponding to the exact LO SM amplitude~${\cal M}_{\rm exact}^{(0)}$ for a specific partonic~process~$X$. 

To compute a differential hadronic cross section, it is necessary to sum over the various partonic processes, convolute them with the corresponding parton luminosities, and integrate over the relevant phase space. In doing so, the numerator and denominator of~(\ref{eq:C1}) must be treated separately, and the corresponding ratio of the integrated quantities should then be formed. In our case, this is achieved by incorporating the formulas~(\ref{eq:regulated}) into the MC code developed in~\cite{Niggetiedt:2024nmp}. This generator allows for the simulation of~ggF Higgs production within the full SM, achieving next-to-next-to-leading order plus parton shower~(NNLO+PS) accuracy in QCD, using the {\tt POWHEG BOX RES}~\cite{Jezo:2015aia} framework. The interpolation of the grids describing the kinematic dependence of the regulated squared matrix elements is performed using cubic B-splines, implemented via the {\tt bspline-fortran} code~\cite{bspline-fortran}. We adopt the following input parameters: $m_h = 125 \, {\rm GeV}$, $m_t = 173 \, {\rm GeV}$, and $G_F = 1.166379 \cdot 10^{-5} \, {\rm GeV}^{-2}$, the latter of which leads to $v = 2^{-1/4} \hspace{0.5mm} G_F^{-1/2} = 246.22 \, {\rm GeV}$. The center-of-mass energy of the proton-proton collisions is taken to be $\sqrt{s} = 13.6 \, {\rm TeV}$, and the parton distribution functions used in our MC simulations are {\tt NNPDF40\_nlo\_as\_0118}~\cite{NNPDF:2021njg} with $\alpha_s (M_Z) = 0.1180$, resulting in $\alpha_s(m_h) = 0.1127$. The factorization and renormalization scales are both set to $\mu_F = \mu_R = m_h$. Finally, we impose a cut of $p_{T,h} > 20 \, {\rm GeV}$ to obtain finite predictions for the total $pp \to hj$ cross section. 

\begin{figure}[!t]
\begin{center}
\includegraphics[width=0.65\textwidth]{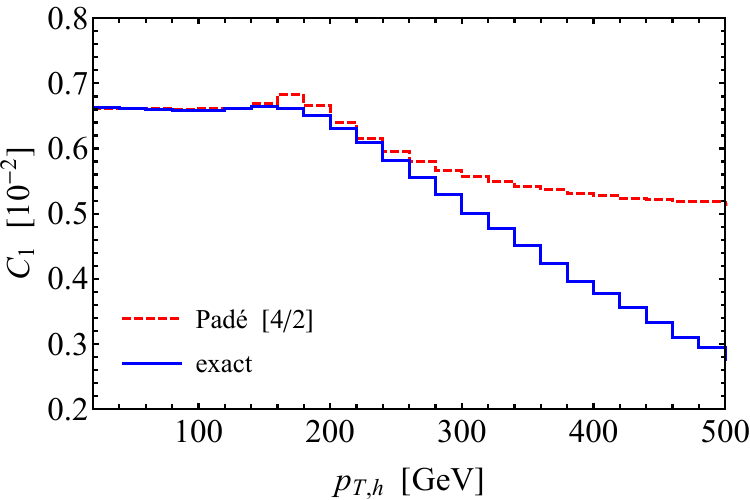} 
\vspace{2mm}
\caption{\label{fig:pThspectrum} The non-universal ${\cal O} (\bar c_6)$ corrections to the~$p_{T,h}$ spectrum in $pp \to jh$ production, as defined in~(\ref{eq:C1}), are illustrated. The blue solid line represents our prediction for $C_1$, which is based on the exact computation of the relevant squared matrix elements using~(\ref{eq:regulated}). For comparison, the best prediction for the correction $C_1$ based on a HTE and a Pad{\'e} approximation, as taken from~\cite{Gao:2023bll}, is also shown as a red dashed line.}
\end{center}
\end{figure}

In~Figure~\ref{fig:pThspectrum}, we present our results for the non-universal ${\cal O} (\bar c_6)$ corrections to the~$p_{T,h}$ spectrum in Higgs plus jet production, as defined in~(\ref{eq:C1}). We first notice that~the correction~$C_1$ approaches a constant value at low $p_{T,h}$. As discussed in~\cite{Gorbahn:2019lwq,Gao:2023bll}, when~$p_{T, h} \to 0$, the non-universal ${\cal O} (\bar c_6)$ corrections to $pp \to hj$ production in~ggF converge to those for inclusive $gg \to h$ production. Based on our chosen input parameters, we obtain the non-universal ${\cal O} (\bar c_6)$ corrections to the inclusive Higgs production cross section in the~ggF~channel as $C_1 = 0.684 \cdot 10^{-2}$. This value updates the previous numerical results reported in~\cite{Degrassi:2016wml,Gorbahn:2019lwq,Gao:2023bll}. Additionally, we observe that the non-universal correction $C_1$ decreases with~$p_{T,h}$ and tends to zero as $p_{T,h} \to \infty$. In this limit, only the Higgs WFR contributes to the $p_{T,h}$ spectrum in $pp \to hj$ production. A similar decoupling behavior was observed in the context of the non-universal~${\cal O} (\bar c_6)$ corrections to $pp \to Vh$ production~in~\cite{Bizon:2016wgr}. For comparison, we also include in~Figure~\ref{fig:pThspectrum} results from~\cite{Gao:2023bll}, which employs a HTE with a Pad{\'e} approximation to extend the convergence range of the~HTE. The plot clearly shows that while the Pad{\'e} approximation accurately represents $C_1$ at low~$p_{T,h}$, it~begins to deviate near the top-quark threshold and overestimates the magnitude of the non-universal~${\cal O} (\bar c_6)$ corrections at high~$p_{T,h}$. This~finding strongly suggests that, in general, an accurate description of the~$C_1$ contributions to the $pp \to hj$ distributions in the momentum transfer regime above the top-quark threshold requires a calculation that fully accounts for the dependence of the squared matrix elements on the Higgs and top-quark masses throughout the entire~$2 \to 2$ phase~space.

\section{Conclusion}
\label{sec:conclusions}

In this article, we have computed the non-universal~${\cal O} (\bar c_6)$ corrections to the two-loop amplitudes for $gg \to hg$, $qg \to hq$, and $q \bar q \to hg$. These contributions are essential for studying the precise impact of a modified cubic Higgs self-coupling on the differential Higgs plus jet production cross section. Unlike earlier calculations~\cite{Gorbahn:2019lwq,Gao:2023bll}, which are reliable only below the top-quark threshold, our computation is expected to deliver accurate numerical results across the entire $2 \to 2$ phase space. Our calculation employs the methodology described and successfully applied in the publications~\cite{Czakon:2020vql,Czakon:2021yub,Czakon:2023kqm,Niggetiedt:2024nmp,Czakon:2024ywb}. It is based on cutting-edge multi-loop techniques, such as deep HTEs, integral reductions to master integrals, and the application of differential equations, to facilitate efficient numerical evaluation of the relevant squared matrix elements. Using our computational setup, we created dense two-dimensional grids for a fixed value of $x = m_h^2/m_t^2 = 12/23$, corresponding to physical Higgs and top-quark masses of $m_h \simeq125 \, {\rm GeV}$ and $m_t \simeq 173 \, {\rm GeV}$, respectively. These grids are ideal for interpolating the relevant squared matrix elements for the $gg \to hg$, $q g \to hq$, and $q \bar q \to hg$ processes over the entire $2 \to 2$ phase space. Each grid contains close to $7 \cdot 10^6$ high-precision phase-space points. We~believe that these grids will also be useful for phenomenological applications involving slightly different fixed ratios $x = m_h^2/m_t^2$, closely matching the actual ratio with an accuracy of better than $1\%$. The~numerical grids are available from the authors upon request.

Using the calculated two-dimensional grids for the squared matrix elements, we have computed the non-universal~${\cal O} (\bar c_6)$ corrections to the $p_{T,h}$ distribution in $pp \to hj$ production for arbitrary values of $p_{T,h}$. We first observed that the corrections level off to a constant value at low $p_{T,h}$, consistent with the value of $C_1$ found for the inclusive Higgs production cross section in $gg \to h$. As discussed in~\cite{Gorbahn:2019lwq,Gao:2023bll}, this behavior is expected. Using our input parameters, we calculated the non-universal ${\cal O} (\bar c_6)$ corrections to the inclusive Higgs production cross section in the ggF channel to be $C_1 = 0.684 \cdot 10^{-2}$, which updates the previous numerical results reported in~\cite{Degrassi:2016wml,Gorbahn:2019lwq,Gao:2023bll}. Furthermore, we noted that the non-universal correction $C_1$ decreases with increasing~$p_{T,h}$ and approaches zero as $p_{T,h} \to \infty$. In this limit, only the Higgs WFR contributes to the $p_{T,h}$ spectrum in $pp \to hj$ production. We~also compared our results with those from~\cite{Gao:2023bll}, which uses a HTE with a Pad{\'e} approximation to extend the convergence range of the~HTE. Our~comparison demonstrated that, in general, only a calculation that accounts for the exact dependence of the squared matrix elements on the Higgs and top-quark masses across the entire $2 \to 2$ phase space can accurately describe the relevant $pp \to hj$ distributions in the momentum transfer regime above the top-quark threshold. 

With the exact squared matrix elements derived in this work, it is now possible to compute the loop-induced effects involving $\bar c_6$ to the relevant differential contributions in the whole phase space not only in the case of the $pp \to hjj$~\cite{Degrassi:2016wml, Bizon:2016wgr}, $pp \to Vh$~\cite{Degrassi:2016wml,Bizon:2016wgr}, $p p \to t \bar t h$~\cite{Degrassi:2016wml}, and $p p \to t h j$~\cite{Maltoni:2017ims} process, but finally also for $pp \to hj$ production. The phenomenological implications of our novel results will be studied elsewhere. In~particular, a detailed analysis of the prospects of future LHC runs to constrain the Wilson coefficient~$\bar c_6$ using differential information in Higgs plus jet events will be presented. The inclusion of the exact two-loop amplitudes for $gg \to hg$, $qg \to hq$, and $q \bar q \to hg$ with a modified cubic Higgs self-coupling into the NNLO+PS QCD generator for ggF Higgs production within the full~SM~\cite{Niggetiedt:2024nmp} will be crucial in the context of this forthcoming study.

\acknowledgments{We are grateful to Li Lin Yang and the co-authors of~\cite{Gao:2023bll} for providing us with their results in electronic form for the Pad{\'e} approximations of the non-universal~${\cal O} (\bar c_6)$ corrections to the $p_{T,h}$ distribution in $pp \to hj$ production. A special thanks also to Jakob Linder, Marius Wiesemann, and, last but certainly not least, Giulia Zanderighi for their help with {\tt POWHEG BOX RES}. The Feynman diagrams shown in this article have been drawn by means of {\tt FeynGame}~\cite{Harlander:2020cyh,Harlander:2024qbn}.}



%

\end{document}